\documentclass[
superscriptaddress,
twocolumn,
amsmath,amssymb,
longbibliography,
aps,
prl,
]{revtex4-1}
\usepackage{graphicx}
\usepackage{dcolumn}
\usepackage{bm}
\usepackage{color}
\usepackage{hyperref}
\usepackage{xcolor}

\newcommand{\br}[1]{\pmb{\mathrm{#1}}}

\begin{document}


\title{Efficient generation of extreme terahertz harmonics in 3D Dirac Semimetals}

\author{J. Lim}
\affiliation{%
Science, Math and Technology, Singapore University of Technology and Design,  Singapore
}%

\author{Y. S. Ang}
\email{yeesin_ang@sutd.edu.sg}
\affiliation{%
Science, Math and Technology, Singapore University of Technology and Design,  Singapore
}%

\author{F. J. Garc\'{i}a de Abajo}
\affiliation{%
ICFO-Institut de Ciencies Fotoniques, The Barcelona Institute of Science and Technology, Barcelona, Spain.
}%
\affiliation{ICREA-Instituci\'{o} Catalana de Recerca i Estudis Avancats, Barcelona, Spain}

\author{I. Kaminer}
\affiliation{%
Department of Electrical Engineering, Technion, Haifa, Israel
}%

\author{L. K. Ang}
\email{ricky_ang@sutd.edu.sg}
\affiliation{%
Science, Math and Technology, Singapore University of Technology and Design, Singapore
}%

\author{L. J. Wong}
\email{liangjie.wong@ntu.edu.sg}
\affiliation{
School of Electrical and Electronic Engineering, Nanyang Technological University, Singapore
}

\date{\today}


\begin{abstract}
Frequency multiplication of terahertz signals on a solid state platform is highly sought-after for the next generation of high-speed electronics and the creation of frequency combs. Solutions to efficiently generate extreme harmonics (up to the $31^{\rm{st}}$ harmonic and beyond) of a terahertz signal with modest input intensities, however, remain elusive. Using fully nonperturbative simulations and complementary analytical theory, we show that 3D Dirac semimetals (DSMs) have enormous potential as compact sources of extreme terahertz harmonics, achieving energy conversion efficiencies beyond $10^{-5}$ at the $31^{\rm{st}}$ harmonic with input intensities on the order of $10$ MW/cm$^2$, over $10^5$ times lower than in conventional THz high harmonic generation systems. Our theory also reveals a fundamental feature in the nonlinear optics of 3D DSMs: a distinctive regime where higher-order optical nonlinearity vanishes, arising as a direct result of the extra dimensionality in 3D DSMs compared to 2D DSMs. Our findings should pave the way to the development of efficient platforms for high-frequency terahertz light sources and optoelectronics based on 3D DSMs.
\end{abstract}

\pacs{Valid PACS appear here}
\maketitle
High-harmonic generation (HHG) is a nonlinear process involving the emission of light at integer multiples of the driving laser frequency. HHG in gaseous media is a well-established method of generating high-frequency light and broadband frequency combs, which provide attosecond resolution in the study of atomic and materials phenomena ~\cite{Itatani2004,Worner2010,Goulielmakis2010,Calegari2014}.  More recently, HHG has been demonstrated in solids~\cite{Ghimire2011HHG,Schubert2014,Luu2015,Ndabashimiye2015, You2017,You2017_2,UzanNatPhoton2020},  revealing its potential for realizing chip-integrable optoelectronics and compact radiation sources.  However, solid-state HHG typically requires strong driving fields exceeding $1$ GV/m (i.e., on the order of TW/cm$^{2}$ peak power)~\cite{Ghimire2011HHG,Schubert2014,Luu2015,Ndabashimiye2015, You2017,You2017_2,UzanNatPhoton2020,Ghimire2019}. With rapidly growing interest in the terahertz (THz) regime as the answer to next-generation computation, imaging and communications, the ability to perform efficient THz HHG using modest input intensities on a solid-state platform is a highly sought-after solution.

\begin{figure*}[ht!]
\centering
\includegraphics[width = 180mm]{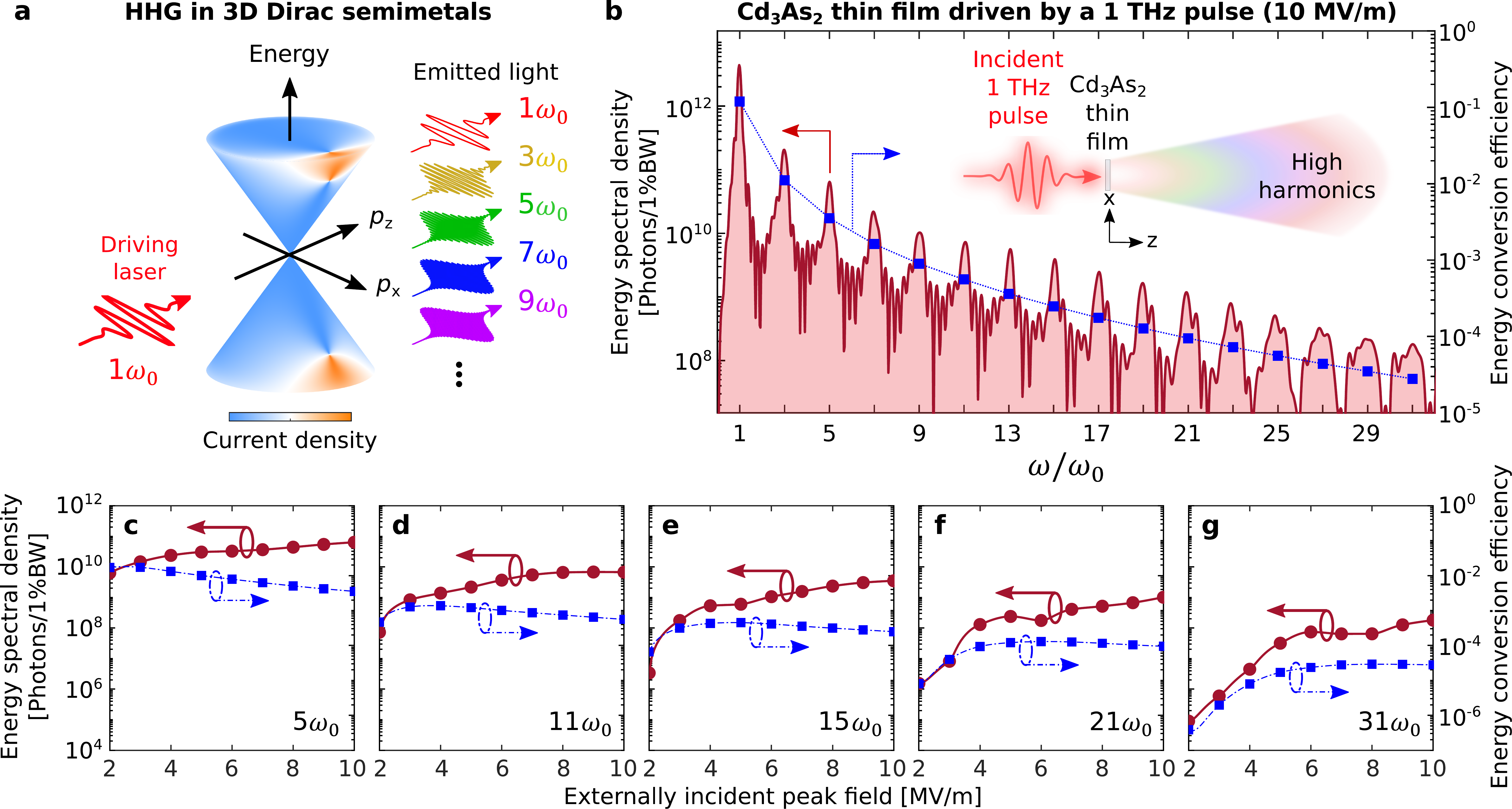}
\caption{Highly efficient generation of extreme harmonics (up to $31^{\rm{st}}$ harmonic) in 3D DSM $\mathrm{Cd_{3}As_{2}}$ at modest driving field strengths. (a) HHG in a 3D DSM occurs when a driving laser pulse (red-arrowed waveform) induces carrier oscillations (current density depicted at one instant in time on the Dirac cone graphic) and transitions which lead to the emission of high harmonic light (multi-colored output arrows). Driving a $\mathrm{Cd_{3}As_{2}}$ thin film with a linearly polarized pulse of central frequency 1 THz and peak field strength 10 MV/m  ((b) inset) produces the emitted spectrum shown in (b), where we see that harmonics up to the $31^{\rm{st}}$ order and beyond can be generated at energy conversion efficiencies well in excess of $10^{-5}$. (c)-(g) show the change in the output energy spectrum and conversion efficiency as a function of externally incident driving field strength. These results reveal that the generation of extreme harmonics continue to remain relatively efficient even at much lower field strengths. The markers in (c)-(g) are the result of our numerical simulations, whereas the connecting lines are visual guides.  We considered a 250 nm thick $\mathrm{Cd_{3}As_{2}}$ thin film of radius 1 mm and
 Fermi energy $\mathcal{E}_{\rm{F}} = 60$ meV, uniformly illuminated by a 2 ps long pulse of 1 THz peak frequency.}
\label{fig_1}
\end{figure*}

Here, we show that the recently discovered class of 3D Dirac semimetal (DSM) materials is a promising candidate for solid-state THz HHG, achieving high efficiencies with input field intensities on the order of $10$ MW/cm$^2$, more than $10^{5}$ times weaker than required for conventional THz HHG. This applies not only to the generation of lower-order harmonics~\cite{ChingHuaNodalHHG,Cd3As2THzHHG,APLphotonics_YeeSin,Kovalev2019,2020_Ooi_ThzDiracPlasmonics}, but also to the generation of extreme harmonics, up to the $31^{\rm{st}}$ order and beyond.  In particular, by considering a $\mathrm{Cd_{3}As_{2}}$ thin film of 250 nm thickness (readily obtained using methods like molecular beam epitaxy~\cite{Cd3As2THzHHG, PhysRevB.93.121202, Yuan2017, Nishihaya2019a, Nishihayaeaar5668, Liang2017, Kovalev2019}), we predict that conversion efficiencies exceeding $10^{-5}$ can be achieved as high as the $31^{\rm{st}}$ harmonic.  Notably, we find that the $31^{\rm{st}}$ harmonic peak can be brought within 5 orders of magnitude of the driving harmonic using an input field of 10 MV/m and below, well within reach of existing table-top THz sources~\cite{single_cycle_1THz, OR_organic_crystals, HuangOptLett2013, Fulop_mJ_THz,Fulop_THz_OR}. In contrast, conventional solid-state THz HHG requires driving fields approaching 10 GV/m -- both in theory and in experiment -- just to bring the $22^{\mathrm{nd}}$ harmonic peak within 8 orders of magnitude of the driving harmonic~\cite{Schubert2014}.

Our studies also reveal a novel regime where the higher order nonlinearities of the intraband current in 3D DSMs completely vanish. As a result, the efficiency of HHG beyond the $3^{\mathrm{rd}}$ harmonic is greatly suppressed in this regime. As we show, this is a fundamental feature of 3D DSMs directly arising from their extra dimensionality compared to 2D DSMs, which have no such regime of suppression. This breaks the common notion that 2D DSMs -- well known for their nonlinear optical response~\cite{Wright2009,Shareef2012,Yoshikawa736, Cox2017, THzHHG_hotcarriers, PhysRevLett.122.053901,Giorgianni2016} -- share the same essential physics as 3D DSMs. We identify a physical quantity -- termed the critical field strength -- that divides the regime of nonlinear suppression from the regime of extreme nonlinearity (and hence highly efficient HHG) in 3D DSMs. We show that in the latter regime, 3D DSMs have orders-of-magnitude performance enhancements compared to existing solid-state HHG platforms.


\textbf{RESULTS. } A laser pulse impinging on a DSM material induces carrier oscillations within (intraband current) and carrier transitions between (interband current) the upper and lower bands of the Dirac cone band structure. These carrier oscillations and transitions in turn emit light peaked at multiples of the driving laser frequency (Fig. 1a).

The scenario we consider is illustrated in the inset of Fig. 1b: An external laser pulse is normally incident on a DSM material, which emits multiple harmonics of the incident pulse. Maxwell's equations are used to model the electrodynamics of the pulse and its interaction with the DSM material. The properties of the DSM material are in turned determined using nonpertubative time-domain quantum simulations. Full details of the treatment are given in Methods, with the key points as follows.

To model the dynamics in time $t$ of an electron in a Dirac cone band structure, we start with the effective Hamiltonian
\begin{equation}
\mathrm{i}\hbar\frac{\partial}{\partial t} = \sum_{j}v_{j}\sigma_{j}p_{j}
\label{eqn_4band_TDDE}
\end{equation}
where $\hbar$ is the reduced Planck constant, $v_{j}$ are the Fermi velocities along Cartesian directions $j$,  $\sigma_{j}$ are the Pauli matrices, and $p_{j}$ are the initial electron momenta, with $j \in \{x,y,z\}$ and $j\in\{x,y\}$  for 3D and 2D DSMs, respectively. We include the field via a modified minimal coupling substitution $\pmb{\mathrm{p}} \rightarrow \pmb{\pi}(t) = \pmb{\mathrm{p}}+e\pmb{\mathrm{a}}(t)$, where $e$ is the elementary charge, and $\pmb{\mathrm{a}}(t)$ is the modified vector potential inside the DSM, which is related to the electric field $\br{E}(t)$ by the relation $\br{a}(t) = -e^{-t/\tau}\int_{-\infty}^{t}\br{E}(t')e^{t'/\tau}dt'
$~\cite{PhysRevB.95.125408}, where the inelastic intraband scattering time is $\tau$. In the absence of intraband scattering, i.e., $\tau\rightarrow\infty$, $\br{a}(t)$ is exactly the vector potential $\br{A}(t)$. From Eq. (\ref{eqn_4band_TDDE}), we obtain the total induced current $J_{i}(t)$, in terms of  the population inversion $\mathcal{N}_{\pmb{\mathrm{p}}}(t)$ and the interband coherence $\mathit{\Gamma}_{\pmb{\mathrm{p}}}(t)$, as
\begin{equation}
\begin{split}
J_{i} &= \frac{gev_{i}}{(2\pi\hbar)^{n}}\int  \Bigg{\{}\Big{(}\mathcal{N}_{\pmb{\mathrm{p}}}+1\Big{)}\frac{v_{i}\pi_{i}}{\mathcal{E}} - 2\mathit{\Lambda}_{i}\mathrm{Re}\Big{(}\mathit{\Gamma}_{\pmb{\mathrm{p}}}\Big{)} \\
&\qquad\qquad\qquad\qquad + 2\mathit{\Delta}_{i}\mathrm{Im}\Big{(}\mathit{\Gamma}_{\pmb{\mathrm{p}}}\Big{)}\Bigg{\}}d^{n}\pmb{\mathrm{p}},
\end{split}
\label{eqn_arbitrary_J}
\end{equation}
where $g = 4$ is the combined valley and spin degeneracy and the integral extends over all momentum space. For 3D DSMs, we take $n = 3$ and $i\in\{x,y,z\}$; for 2D DSMs, we have  $n = 2$ and $i\in\{x,y\}$. We also define the instantaneous energy $\mathcal{E}(t) = \sqrt{\sum_{i}v_{i}^{2}\pi_{i}^{2}(t)}$, $(\mathit{\Lambda}_{x},\mathit{\Lambda}_{y},\mathit{\Lambda}_{z}) = (\cos\theta\cos\phi,\cos\theta\sin\phi,-\sin\theta)$, and $(\mathit{\Delta}_{x},\mathit{\Delta}_{y},\mathit{\Delta}_{z}) = (-\sin\phi,\cos\phi,0)$, where $\theta(t) = \arccos[v_{z}\pi_{z}(t)/\mathcal{E}(t)]$, and $\phi(t) = \mathrm{arctan}[v_{y}\pi_{y}(t)/v_{x}\pi_{x}(t)]$.  For 2D DSMs, we set $\theta = \pi/2$ and $J_{z} = v_{z} = 0$, obtaining an expression that reduces to the special case of  isotropic 2D DSMs in~\cite{PhysRevB.95.125408, PhysRevB.82.201402, 1367-2630-15-5-055021, Cox2017,Chizhova2017} when we further set $v_{x} = v_{y} = v_{\rm{F}}$.  In Eq. (\ref{eqn_arbitrary_J}), the first term of the integrand represents the contribution of intraband current. The remaining second and third  terms of the integrand represent the contribution of interband  current. The effect of finite temperature and interband carrier scattering are taken in account in the solutions to $\mathcal{N}_{\br{p}}(t)$ and $\mathit{\Gamma}_{\br{p}}(t)$.

We evaluate Eq. (\ref{eqn_arbitrary_J}) to obtain a fully-closed-form, nonperturbative expression for the total current that is valid when the intraband current dominates the material response, as is the case in our regimes of interest.  In 3D DSMs, the form of the total current depends on the value of the parameter $e\Phi(t)/\mathcal{E}_{\rm{F}}$, where $\Phi(t) = \sqrt{\sum_{i}v_{i}^{2}A_{i}^{2}(t)}$, $i\in\{x,y,z\}$, and $\mathcal{E}_{\rm{F}}$ is the Fermi level. We refer to the field strength that exactly satisfies $e\Phi/\mathcal{E}_{\rm{F}} = 1$ as the critical field strength. When $e\Phi(t)/\mathcal{E}_{\rm{F}} < 1$, we obtain the $x$-component of the current as
\begin{equation}
J_{x}^{\mathrm{3D,sub}}(t) = -\frac{ge^{2}v_{x}}{6\pi^{2}\hbar^{3}v_{y}v_{z}}A_{x}(t)\Bigg{[}\mathcal{E}_{\rm{F}}^{2} - \frac{e^{2}}{5}\sum_{i}v_{i}^{2}A_{i}^{2}(t)\Bigg{]},
\label{eqn_3D_intra_arb_pol_sub}
\end{equation}
where $i\in\{x,y,z\}$.  When $e\Phi(t)/\mathcal{E}_{\rm{F}} > 1$, the expression becomes
\begin{equation}	
\begin{split}
J_{x}^{\mathrm{3D,sup}}(t) &= -\frac{ge\mathcal{E}_{\rm{F}}^{3}}{6\pi^{2}\hbar^{3}v_{y}v_{z}}\times \\	&\quad \frac{v_{x}A_{x}(t)}{\sqrt{\sum_{i}v_{i}^{2}A_{i}^{2}(t)}}\Bigg{[}1 - \frac{\mathcal{E}_{\rm{F}}^{2}}{5e^{2}\sum_{i}v_{i}^{2}A_{i}^{2}(t)}\Bigg{]}.
\label{eqn_3D_intra_arb_pol_sup}
\end{split}
\end{equation}
Equations (\ref{eqn_3D_intra_arb_pol_sub}) and (\ref{eqn_3D_intra_arb_pol_sup}) were obtained assuming that the driving frequencies $\omega$ satisfy the relation $\hbar\omega \ll 2\mathcal{E}_{\rm{F}}$ (intraband current dominantes), and that the initial Fermi distribution $f_{\rm{D}}(\mathcal{E})$ is well-approximated by $f_{\rm{D}}(\mathcal{E}) = 1$ when $\mathcal{E} < \mathcal{E}_{\rm{F}}$, and $f_{\rm{D}}(\mathcal{E}) = 0$ when $\mathcal{E} > \mathcal{E}_{\rm{F}}$ (a condition that holds exactly when $T = 0$ K). Details of the derivation are presented in Supplementary Information (SI) Section IA. Throughout this work, we will term $e\Phi_{\mathrm{max}}/\mathcal{E}_{\rm{F}} < 1$ the subcritical regime, and $e\Phi_{\mathrm{max}}/\mathcal{E}_{\rm{F}} > 1$ the supercritical regime, where $\Phi_{\mathrm{max}}$ is the maximum amplitude of $\Phi(t)$. It is noteworthy that that the first- and third-order conductivities of 3D DSMs extracted from Eq.~(\ref{eqn_3D_intra_arb_pol_sub}) agree with known results~\cite{PhysRevB.93.235417,APLphotonics_YeeSin}. At the same time,  we note that the full, non-perturbative response of 3D DSMs has never been presented until now in equations (\ref{eqn_3D_intra_arb_pol_sub}) and (\ref{eqn_3D_intra_arb_pol_sup}).

\begin{figure*}[ht!]
\centering
\includegraphics[width = 150mm]{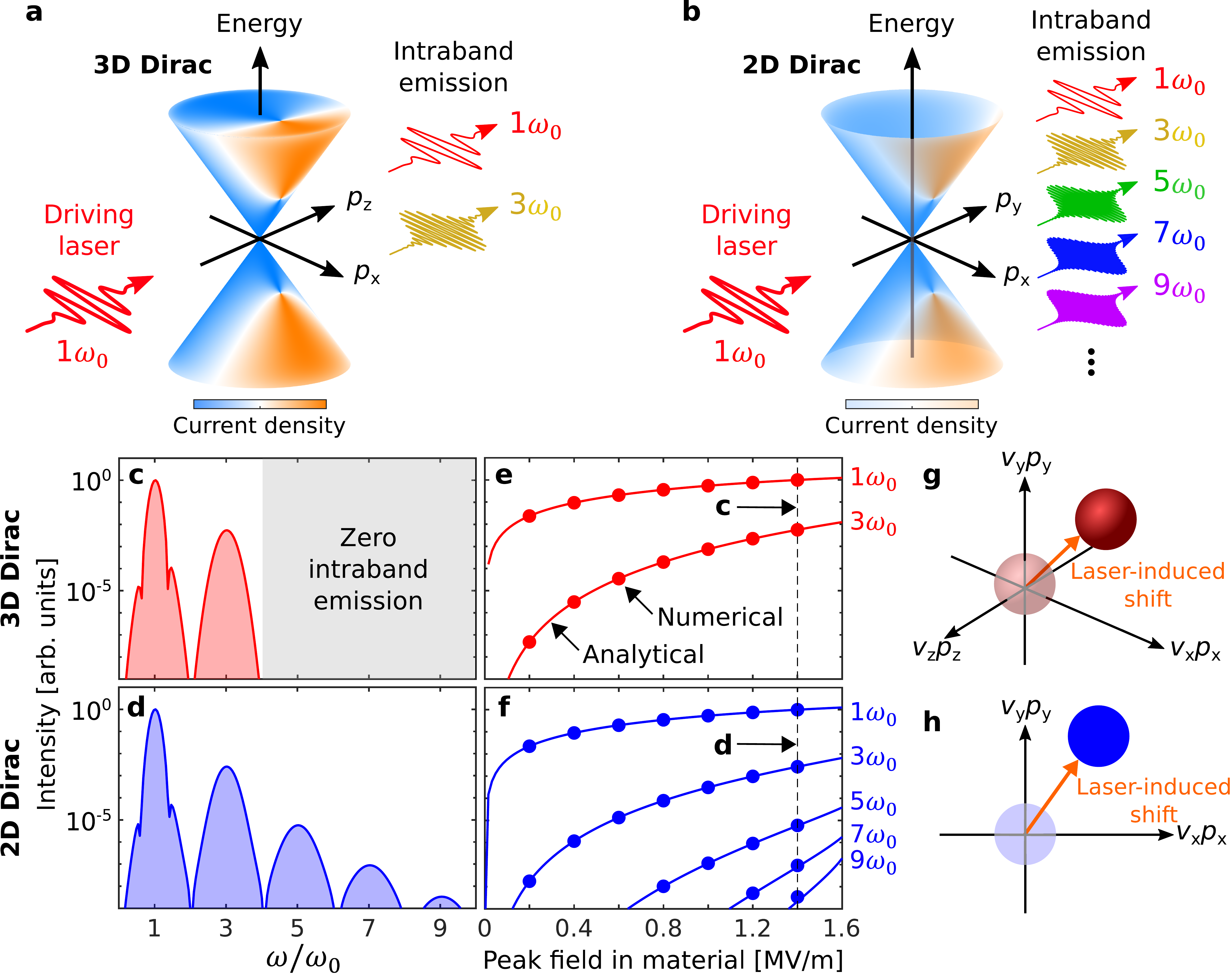}
\caption{Zero higher-order intraband emission from 3D DSMs at subcritical field strengths, leading to strongly suppressed nonlinearities (including HHG) in this regime. Near or below the critical field strength, 3D DSMs (a) cannot efficiently generate harmonics  beyond the third order, unlike 2D DSMs (b). As noted in the text, however, 3D DSMs can achieve energy conversion efficiencies orders of magnitude beyond 2D DSMs and other conventional solid-state HHG platforms. The regime of nonlinear suppression is thus an important fundamental aspect of 3D DSMs, but does not prevent 3D DSMs from emitting high harmonics efficiently under the right conditions (as we see in Fig.~\ref{fig_1}). We show the numerically computed spectra in (c) and (d) for a 3D DSM and a 2D DSM respectively.  As the driving field strength is less than the critical field strength, the intraband current in 3D DSMs contain no nonlinearities beyond the $3^{\mathrm{rd}}$ harmonic, as we see from the output spectrum in (c). In contrast, the output from a 2D DSM (d) contains all harmonics. That this phenomenon holds for all field strengths below the critical field strength is illustrated in (e) and (f). The interband current in 3D DSMs does give rise to output beyond the $3^{\mathrm{rd}}$ harmonic, but its contribution is very weak and has been verified to fall below the range of plotted intensities. In (e) and (f), we also see that the results of our fully closed-form, nonperturbative expressions (solid curves) are in excellent agreement with rigorous numerical simulations (solid circles). The reason for the vanishing of higher-order nonlinearities in 3D DSMs -- although no such phenomena occurs in 2D DSMs -- lies in the extra dimensionality 3D DSMs possess compared to 2D DSMs. Although 3D and 2D DSMs share the same expression for current Eq. (2), the required integration over 3D momentum space (g) for 3D DSMs as opposed to integration over 2D momentum space (h) for 2D DSMs, leads to very different behavior in these materials. For all cases in this figure, we consider Fermi velocities $v_{x} = v_{y} = v_{z} = 10^{6}$ m/s (same as graphene's~\cite{RevModPhysNeto}), Fermi energy $\mathcal{E}_{\mathrm{F}} = 250$ meV at temperature $T= 0$ K, and no carrier scattering.}
\label{fig_2}
\end{figure*}

The HHG energy spectrum presented in Fig.~\ref{fig_1}b indicates that extremely efficient intraband HHG up to the $31^{\rm{st}}$ order and beyond can be generated with significant efficiencies at modest driving laser energies. We consider a 250 nm thick $\mathrm{Cd_{3}As_{2}}$ thin film (readily grown using molecular beam epitaxy ~\cite{Cd3As2THzHHG, PhysRevB.93.121202, Yuan2017, Nishihaya2019a, Nishihayaeaar5668, Liang2017, Kovalev2019}) of Fermi level $\mathcal{E}_{\rm{F}} = 60$ meV, driven by a normally incident, 2 ps-long pulse of peak frequency 1 THz and peak field amplitude $E_{0,x} = 10$ MV/m (Fig.~\ref{fig_1}b inset). We used experimentally verified values
of the Fermi velocities for $\mathrm{Cd_{3}As_{2}}$: $(v_{x},v_{y},v_{z}) = (1.28,1.3,0.33)  \times10^{6}$ m/s~\cite{Liu2014a}.  We find that the $31^{\mathrm{st}}$ harmonic of the output energy spectral density lies within 5 orders of magnitude of the fundamental harmonic. This compares favorably with the performance of conventional solid-state THz HHG, where the $22^{\rm{nd}}$ harmonic lies within 8 orders of magnitude of the fundamental harmonic, when using input intensities $10^5$ times stronger than what we consider~\cite{Schubert2014}.  As we operate in the supercritical regime $e\Phi(t)\gg \mathcal{E}_{\rm{F}}$, Eq. (\ref{eqn_3D_intra_arb_pol_sup}) shows that the intraband current approaches $J_{x}^{\mathrm{3D,sup}}(t) \rightarrow -\mathrm{sgn}[A_{x}(t)]ge\mathcal{E}_{\rm{F}}^{3}/(6\pi^{2}\hbar^{3}v_{y}v_{z})$, which describes a square-wave temporal profile containing only odd-ordered frequency components. In Fig.~\ref{fig_1}, we consider $T = 0$ K and the absence of carrier scattering. As shown in SI Section II, the high output intensities at higher harmonics persist for higher temperatures and non-zero carrier scattering.

Figures~\ref{fig_1}c-\ref{fig_1}g, which show the output intensities of various harmonic orders generated by the same $\mathrm{Cd_{3}As_{2}}$ thin film when the incident peak field strength is varied between $2$ MV/m and $10$ MV/m, reveal that efficient HHG performance at extreme harmonics can still be accessed even at more modest field strengths. We define energy conversion efficiency as $U_{N}/U_{\mathrm{in}}$ where $U_{\mathrm{in}}$ is the incident laser energy (considering only the part of the pulse that interacts with the sample) and $U_{N}$ is the output energy of the $N^{\rm{th}}$ harmonic,  obtained by integrating the energy spectrum over a bandwidth of $\omega_0$ about the harmonic peak at $N\omega_0$. For each harmonic peak, we generally observe a rise followed by a fall in the energy spectral density as driving electric field strengths increase, indicative of the saturation of lower harmonics at larger driving fields.

Our studies also reveal the existence of a novel regime in 3D DSMs where where higher order harmonic emission is instead suppressed. As can be seen directly from Eq.~(\ref{eqn_3D_intra_arb_pol_sup}), the intraband current in the subcritical regime ($e\Phi_{\mathrm{max}}/\mathcal{E}_{\rm{F}} < 1$) is made up purely of the $1^{\rm{st}}$ and $3^{\rm{rd}}$ harmonics. As such, the emission due to the intraband current -- which we term intraband emission -- also consists only of the $1^{\rm{st}}$ and $3^{\rm{rd}}$ harmonics (Fig.~\ref{fig_2}a). This is a noteworthy feature of 3D DSMs, especially since 2D DSMs exhibit no such vanishing of higher order harmonics under any condition (Fig.~\ref{fig_2}b). To further illustrate this, the emission spectra of 3D and 2D DSMs in the subcritical regime is shown in Figs.~\ref{fig_2}c,e and \ref{fig_2}d,f respectively. This difference in 3D and 2D DSM behavior can also be seen by directly comparing the intraband current for 2D DSMs, given by Supplementary Eq. (S22) -- which contains every order of nonlinearity -- in SI Section IB with the intraband current for 3D DSMs in Eq. (\ref{eqn_3D_intra_arb_pol_sup}).

The reason for the vanishing of higher-order nonlinearities in 3D DSMs lies in the extra dimensionality 3D DSMs possess compared to 2D DSMs. Due to the extra dimensionality, the integral in Eq. (\ref{eqn_arbitrary_J}) takes place over 3 dimensions in momentum space for 3D DSMs -- resulting in the spherical region of integration shown in Fig.~\ref{fig_2}g -- as opposed to just 2 dimensions in momentum space for 2D DSMs (Fig.~\ref{fig_2}h). It is also possible to mathematically illustrate this vanishing of all higher orders in the 3D DSM subcritical regime using a Legendre polynomial expansion, which we do in SI Section III.

In Fig.~\ref{fig_2}, we consider temperature $T = 0$ K and the absence of carrier scattering. However, we find that even when finite temperatures and carrier scattering effects are considered, the suppression of higher-order light emission remains significant (SI Section IV).  Furthermore, we note that this suppression of HHG in the subcritical regime of 3D DSMs is a phenomenon that is robust against variations in driving field polarization and phase, and Fermi velocity anisotropy (SI Section V).

\begin{figure}[ht]
\centering
\includegraphics[width = 85mm]{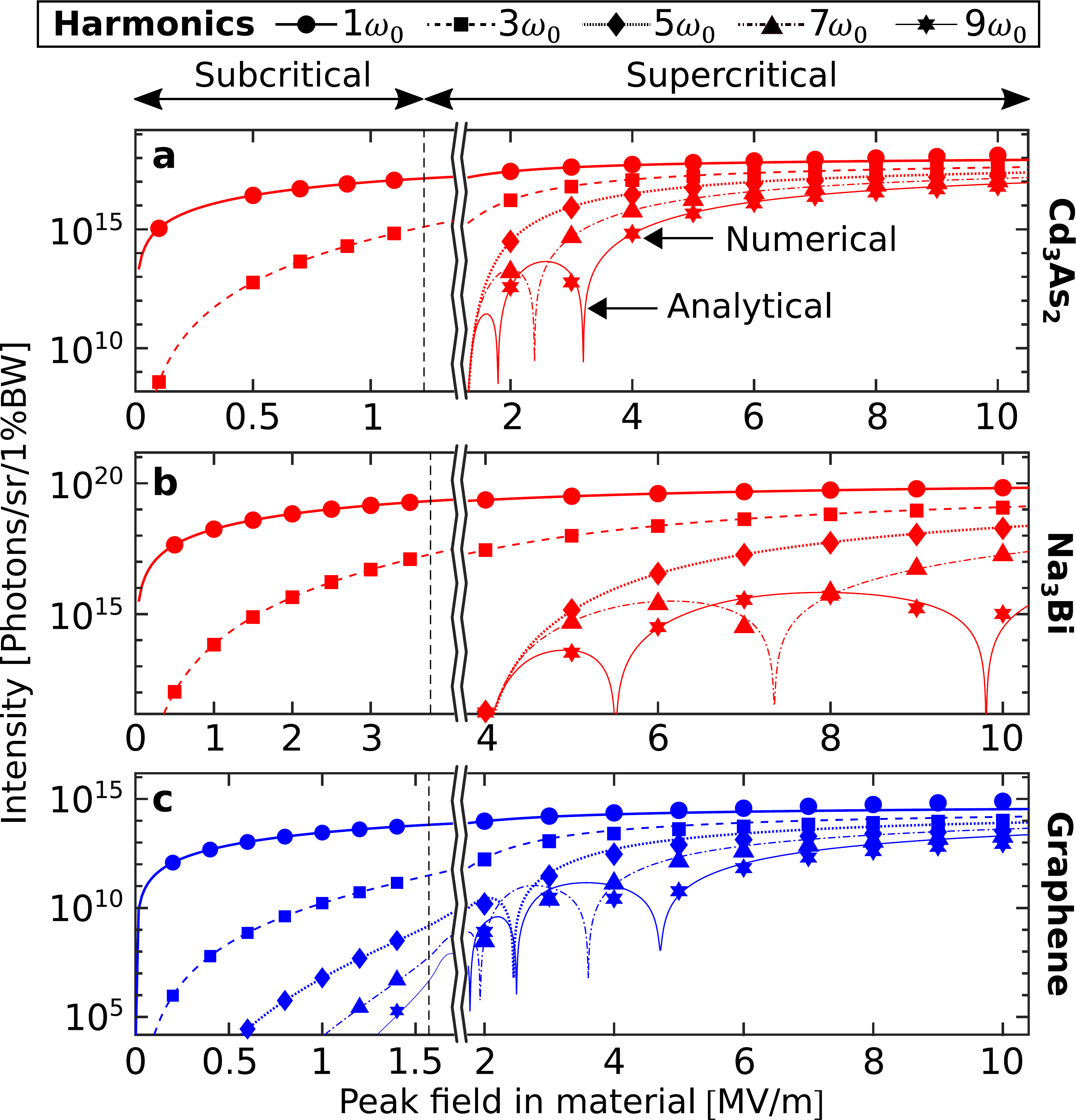}
\caption{High-harmonic generation in 3D DSMs $\mathrm{Cd_{3}As_{2}}$ (a) and $\mathrm{Na_{3}Bi}$ (b), and in 2D DSM graphene (c). Below the critical field strength (vertical black dashed line), the emitted harmonics beyond the $3^{\mathrm{rd}}$ harmonic are greatly suppressed in 3D DSMs. No such suppression occurs in the case of 2D DSMs. Above the critical field strength, however, the intensity of the emitted harmonics in 3D DSMs rapidly increase with increasing field strength. In each panel, markers and lines denote numerical results (which account for both intraband and interband emission) and analytical expressions (which account only for intraband emission) respectively. The good agreement indicates the dominance of intraband emission in our regime of interest. For the sake of clarity, we plot only up to the ninth harmonic. }
\label{fig_3}
\end{figure}

In Fig.~\ref{fig_3}, we explore HHG in both subcritical and supercritical regimes for experimentally realized 3D DSMs $\mathrm{Cd_{3}As_{2}}$~\cite{Liu2014a} and $\mathrm{Na_{3}Bi}$~\cite{Science_Na3Bi_discovery}. We used experimentally obtained Fermi velocities for $\mathrm{Na_{3}Bi}$: $(v_{x},v_{y},v_{z}) = (4.17,3.63,0.95)\times10^{5}$ m/s~\cite{Science_Na3Bi_discovery}, and, unless otherwise specified, the same parameters as in Fig.~\ref{fig_2}. Once again, we observe excellent agreement between our numerical spectra (markers) and closed-form expressions given by equations (\ref{eqn_3D_intra_arb_pol_sub}) and (\ref{eqn_3D_intra_arb_pol_sup}) (curves). When driven by fields larger than the critical field strength,  we observe that 3D DSMs behave in a qualitatively similar manner as graphene~\cite{PhysRevB.95.125408, PhysRevB.82.201402, PhysRevLett.105.097401}. Note, however, that a comparison between $\mathrm{Cd_{3}As_{2}}$ and (single-layer) graphene shows that $\mathrm{Cd_{3}As_{2}}$ can surpass the power output of graphene by over 2 orders of magnitude under the same input conditions, especially at higher harmonics. Using a hypothetical multilayer graphene-dielectric structure to achieve performance comparable with $\mathrm{Cd_{3}As_{2}}$ would require at least 9 layers of graphene-dielectric, a structure that has yet to be realized (SI Section VI). Our results show that DSMs with higher Fermi velocities along the direction of the driving field polarization are able to access the supercritical regime at lower field strengths, as expected from the fact that the critical field  strength scales proportionally with $\Phi = \sqrt{\sum_{i}v_{i}^{2}A_{i}^{2}}$. As a result, we see in Figs.~\ref{fig_3}a-\ref{fig_3}c that higher harmonics rapidly emerge at a lower field strength for $\mathrm{Cd_{3}As_{2}}$ as compared to $\mathrm{Na_{3}Bi}$ and graphene since $\mathrm{Cd_{3}As_{2}}$ has the highest Fermi velocity in the direction of the driving laser polarization.


%

\textbf{DISCUSSION.} Our work reveals two distinct operation regimes in 3D DSMs: The supercritical regime where optical nonlinearities are strong and HHG efficiently generates harmonics up to the $31^{\rm{st}}$ order and beyond; and the subcritical regime where nonlinearities beyond the $3^{\rm{rd}}$ order completely vanish from the intraband current, resulting in greatly diminished higher harmonic output. While much work has risen around 3D DSMs as bulk versions of 2D DSMs~\cite{Cd3As2THzHHG, Zhu2017, Wang2017, Meng2018}, our findings break the common notion that these two systems share the same essential physics. As we show, the extra dimensionality in 3D DSMs could lead to much weaker nonlinear response from 3D DSMs in the subcritical regime compared to 2D DSMs, even though 3D DSMs seem to have a larger interaction volume. Choosing parameters that put us in the supercritical regime, however, we see that 3D DSMs can efficiently generate HHG up to the $31^{\rm{st}}$ harmonic and more, well beyond recent experiments that demonstrated frequency upconversion with 3D DSMs up to the $5^{\rm{th}}$~\cite{Cd3As2THzHHG} and $7^{\rm{th}}$ harmonics~\cite{Kovalev2019}. Furthermore, the extreme THz HHG we study is performed at modest driving intensities, over $10^5$ times lower than in conventional THz high harmonic generation systems.

Our studies also reveal the important role of anisotropy in HHG from 3D DSMs. Because the critical field strength scales as $\sqrt{\sum_{i}v_{i}^{2}A_{i}^{2}}$, and the amplitude of the intraband current is proportional to $1/v_{y}v_{z}$ for an $x$-polarized input field in the deep supercritical regime, we see that the value of the Fermi velocity $\br{v}$ has significant influence over the nonlinear optical properties of 3D DSMs. This motivates the development of bandstructure engineering methods that can give us greater control over the Fermi velocity of a 3D Dirac cone, for instance, to enhance the HHG intensity through an appropriate choice of Fermi velocity (as we show through an example in SI Section VII).

Our findings are also relevant to the broader study of 3D DSMs, even beyond high harmonic generation. In particular, we note that the full, non-perturbative response of 3D DSMs has never been presented until now in Eqs. (\ref{eqn_3D_intra_arb_pol_sub})-(\ref{eqn_3D_intra_arb_pol_sup}), where we present them in fully closed-form, analytical expressions. These analytical expressions also allow very convenient implementation in numerical electrodynamics solvers to study the electromagnetic response of 3D DSMs, as we do using an FDTD algorithm. With growing interest in the optical properties of unconventional topological bandstructures~\cite{ChingHuaNodalHHG, YanWeylNodalLine, WeylNematollahi2019, RubioFloquetWeyl}, the theory and numerical implementation we present here could also potentially prove useful in studying the non-perturbative light-driven dynamics of low-energy electrons in various systems.

In summary,  we show the ability of 3D DSMs to efficiently generate extreme THz harmonics up to the $31^{\rm{st}}$ harmonic and beyond, using modest laser intensities on the order of $10$ MW/cm$^{2}$, which are $10^{5}$ times weaker than in conventional solid-state HHG~\cite{Ghimire2011HHG,Schubert2014,Luu2015,Ndabashimiye2015, You2017,You2017_2}.  Our studies reveal two distinct operation regimes for 3D DSMs: the supercritical regime, where the extreme nonlinearity of the 3D DSM response leads to HHG energy conversion efficiencies exceeding $10^{-5}$ at the $31^{\rm{st}}$ harmonic; and the subcritical regime, where intraband emission vanishes and HHG output is greatly diminished. We further show that the vanishing of intraband emission in the subcritical regime is linked to the extra dimensionality of 3D DSMs compared to 2D DSMs, breaking the common notion that 3D DSMs are simply bulk versions of 2D DSMs. Our studies at different temperatures, scattering times, incident field polarization and material anisotropy show that our conclusions are robust under a broad range of parameters.  Our work fills a vital gap in the understanding of nonlinear physics in 3D DSMs, paving the way to the development of highly efficient, chip-integrable THz light sources and optoelectronics.

\clearpage
\newpage
\section{Methods}
\textbf{Nonperturbative massless electron dynamics.} Here, we derive the ordinary differential equations (ODE) governing the coupled behavior of the interband and intraband electron dynamics in a 3D DSM. Our derivation here follows from main text Eq. (\ref{eqn_4band_TDDE}). Using the minimal coupling substitution, the low-energy Hamiltonian reads
\begin{equation}
\mathrm{i}\hbar\frac{\partial}{\partial t} = \hat{\mathcal{H}}_{\br{p}} = \sum_{j}v_{j}\sigma_{j}\pi_{j}(t),
\label{eqn_TDDE_minimal_coupling}
\end{equation}
where $j\in\{x,y,z\}$ runs over Cartesian directions, $\hbar$ is the reduced Planck's constant, $v_{j}$ are the Fermi velocities associate with direction $j$, $\sigma_{j}$ are the Pauli matrices, $\pi_{j}(t)= p_{j} + ea_{j}(t)$ are the components of the modified minimal coupling momentum, $p_{j}$ is the unperturbed electron momentum, $e$ is the elementary change, $a_{j}(t)$ is defined as~\cite{PhysRevB.95.125408}
\begin{equation}
a_{j}(t) = -e^{-t/\tau}\int^{t}_{-\infty}E_{j}(t')e^{t'/\tau}dt'
\label{eqn_modified_vector_potential}
\end{equation}
where $\tau$ is the inelastic intraband scattering time, and $E_{j}(t)$ the time-varying, spatially-uniform electric field. In the absence of intraband scattering, i.e., $\tau \rightarrow \infty$, $a_{j}(t)$ is exactly equivalent to the unmodified vector potential $A_{j}(t) = -\int_{-\infty}^{t}E_{j}(t')dt'$. The fields should include the self-consistent response of the material, but for simplicity we approximate it as the external field, neglecting induced-field effects. Following previous studies on graphene~\cite{PhysRevB.82.201402,1367-2630-15-5-055021,PhysRevB.95.125408}, we write the normalized, time-dependent eigenstates in the adiabatic limit:
\begin{subequations}
\begin{align}
\psi_{\br{p},\rm{c}}(t) &= 
\begin{Bmatrix}
\cos[\theta(t)/2]e^{-\mathrm{i}\phi(t)/2}\\
\sin[\theta(t)/2]e^{+\mathrm{i}\phi(t)/2}
\end{Bmatrix}e^{-\mathrm{i}\Omega(t)}\\
\psi_{\br{p},\rm{v}}(t) &= 
\begin{Bmatrix}
\sin[\theta(t)/2]e^{-\mathrm{i}\phi(t)/2}\\
-\cos[\theta(t)/2]e^{+\mathrm{i}\phi(t)/2}
\end{Bmatrix}e^{+\mathrm{i}\Omega(t)}.
\end{align}
\end{subequations}
The ``$\rm{c}$'' and ``$\rm{v}$'' subscripts denote the conduction and valence band states respectively. We define $\theta(t) = \arccos[v_{z}\pi_{z}(t)/\mathcal{E}(t)]$ and $\phi(t) = \mathrm{arctan}[v_{y}\pi_{y}(t)/v_{x}\pi_{x}(t)]$. The dynamical phase is $\Omega(t) = \hbar^{-1}\int_{t_{0}}^{t}\mathcal{E}(t')~dt'$, where the instantaneous energy is $\mathcal{E}(t) = \sqrt{v_{x}^{2}\pi_{x}^{2} + v_{y}^{2}\pi_{y}^{2} + v_{z}^{2}\pi_{z}^{2}}$.  The most general wave function that satisfies Eq. (\ref{eqn_TDDE_minimal_coupling}) can be constructed by superposing the above eigenstates:
\begin{equation}
\Psi_{\br{p}}(t) = C_{\br{p},\rm{c}}(t)\psi_{\br{p},\rm{c}}(t) + C_{\br{p},\rm{v}}(t)\psi_{\br{p},\rm{v}}(t).
\label{eqn_ansatz}
\end{equation}
The complex coefficients $C_{\pmb{\mathrm{p}},\rm{c}}(t)$ and $C_{\pmb{\mathrm{p}},\rm{v}}(t)$ describe how the probability of finding the electron in either state evolves. By using Eq.~(\ref{eqn_ansatz}) as an ansatz to the Sch\"{o}dinger Eq., i.e., $\mathrm{i}\hbar\partial_{t}\Psi_{\pmb{\mathrm{p}}}(t) = \hat{\mathcal{H}}_{\pmb{\mathrm{p}}}\Psi_{\pmb{\mathrm{p}}}(t)$, and defining the population inversion (difference of electron population between the valence and conduction bands), $\mathcal{N}_{\pmb{\mathrm{p}}}(t) = \lvert C_{\pmb{\mathrm{p}},\rm{c}}(t)\rvert^{2} - \lvert C_{\pmb{\mathrm{p}},\rm{v}}(t)\rvert^{2}$, and the interband coherence, $\mathit{\Gamma}_{\pmb{\mathrm{p}}}(t) = [C_{\pmb{\mathrm{p}},\mathrm{v}}(t)]^{*}C_{\pmb{\mathrm{p}},\mathrm{c}}(t)e^{-2\mathrm{i}\Omega(t)}$, where $^{*}$ denotes the complex conjugate, we can derive a set of ODEs, which describe how these quantities vary in time:
\begin{subequations}
\begin{align}
\label{eqn_3D_NR}
\begin{split}
\dot{\mathcal{N}}_{\pmb{\mathrm{p}}}(t) =& -\frac{1}{\tau}\Big{[}\mathcal{N}_{\pmb{\mathrm{p}}}(t) - \mathcal{N}_{\pmb{\mathrm{p}}}(t_{0})\Big{]} - 2\dot{\theta}(t)\mathrm{Re}\Big{[} \mathit{\Gamma}_{\pmb{\mathrm{p}}}(t) \Big{]} \\
&+ 2\dot{\phi}(t)\sin\theta(t)\mathrm{Im}\Big{[} \mathit{\Gamma}_{\pmb{\mathrm{p}}}(t) \Big{]}
\end{split}\\
\label{eqn_3D_PR}
\begin{split}
\dot{ \mathit{\Gamma}}_{\pmb{\mathrm{p}}}(t) =&  \bigg{[}\mathrm{i}\dot{\phi}(t)\cos\theta(t) -\frac{1}{\tau} - \frac{2\mathrm{i}\mathcal{E}(t)}{\hbar}\bigg{]}\mathit{\Gamma}_{\pmb{\mathrm{p}}}(t)  \\
& + \bigg{[}\dot{\theta}(t)-\mathrm{i}\dot{\phi}(t)\sin\theta(t) \bigg{]}\frac{\mathcal{N}_{\pmb{\mathrm{p}}}(t)}{2}.
\end{split}
\end{align}
\end{subequations}
Here, inelastic interband damping has been introduced phenomenologically through scattering time $\tau$. While $\tau$ in Eqs.~(\ref{eqn_3D_NR}) and (\ref{eqn_3D_PR}) can be different from $\tau$ in Eq.~(\ref{eqn_modified_vector_potential}),  we chose to use the same value for both scattering times. At some initial time, $t_{0}$, long before the laser pulse interacts with the electrons, we have $\mathcal{N}_{\pmb{\mathrm{p}}}(t_{0}) = f_{\rm{D}}[\mathcal{E}(t_{0})] - f_{\rm{D}}[-\mathcal{E}(t_{0})]$ and  $\Gamma_{\pmb{\mathrm{p}}}(t_{0}) = 0$, where  $f_{\rm{D}}(\mathcal{E}) = \{1+\exp[(\mathcal{E} - \mu(T))/k_{\rm{B}}T]\}^{-1}$ is the Fermi-Dirac distribution for a chemical potential $\mu(T)$ at temperature $T$, and $k_{B}$ is the Boltzmann constant.  Equations (\ref{eqn_3D_NR}) and (\ref{eqn_3D_PR}) describe how a massless electron moves within (intraband) and transitions between (interband) energy bands under the influence of driving fields, unlike the semiclassical Boltzmann equation which only describes intraband motion.  In deriving these ODEs, we have made no approximations beyond a constant scattering time and the massless electron limit. The equivalent equations for the 2D case, which have been successful in describing infinitely-extended graphene interacting with a laser~\cite{PhysRevB.95.125408,PhysRevB.82.201402,1367-2630-15-5-055021}, are obtained by neglecting the $z$-components and setting $\theta = \pi/2$, effectively limiting interactions to within the $p_{x}$-$p_{y}$ plane. 

The induced current due to a single momentum value $\br{p}$ is computed as 
\begin{equation}
\pmb{\mathrm{j}}_{\pmb{\mathrm{p}}}(t) = -e\Psi_{\pmb{\mathrm{p}}}^{\dagger}(t)\nabla_{\pmb{\pi}}\hat{\mathcal{H}}_{\pmb{\mathrm{p}}}\Psi_{\pmb{\mathrm{p}}}(t),
\end{equation}
where $\nabla_{\pmb{\pi}}\hat{\mathcal{H}}_{\pmb{\mathrm{p}}} = (v_{x}\sigma_{x},v_{y}\sigma_{y},v_{z}\sigma_{z})$ is the group velocity operator and the $\dagger$ supercript denotes Hermitian conjugate. The individual current components are
\begin{subequations}
\begin{align}
\label{eq_jkx}
\begin{split}
j_{\pmb{\mathrm{p}},x} &= -ev_{x}\Big{\{} \mathcal{N}_{\pmb{\mathrm{p}}}\sin\theta\cos\phi -2\cos\theta\cos\phi\mathrm{Re}\Big{(} \mathit{\Gamma}_{\pmb{\mathrm{p}}}\Big{)}\\
&~\qquad\qquad - 2\sin\phi\mathrm{Im}\Big{(} \mathit{\Gamma}_{\pmb{\mathrm{p}}}\Big{)}\Big{\}}
\end{split}\\
\label{eq_jky}
\begin{split}
j_{\pmb{\mathrm{p}},y} &= -ev_{y}\Big{\{} \mathcal{N}_{\pmb{\mathrm{p}}}\sin\theta\sin\phi - 2\cos\theta\sin\phi\mathrm{Re}\Big{(} \mathit{\Gamma}_{\pmb{\mathrm{p}}}\Big{)}\\
&~\qquad\qquad + 2\cos\phi\mathrm{Im}\Big{(} \mathit{\Gamma}_{\pmb{\mathrm{p}}}\Big{)}\Big{\}}\end{split}\\
\label{eq_jkz}
j_{\pmb{\mathrm{p}},z} &= -ev_{y}\Big{\{} \mathcal{N}_{\pmb{\mathrm{p}}}\cos\theta +2\sin\theta\mathrm{Re}\Big{(} \mathit{\Gamma}_{\pmb{\mathrm{p}}}\Big{)}\Big{\}}.
\end{align}
\end{subequations}
The total induced current is obtained by integrating over all momentum space:
\begin{equation}
\pmb{\mathrm{J}}(t) = \frac{g}{(2\pi\hbar)^{3}}\iiint \pmb{\mathrm{j}}_{\pmb{\mathrm{p}}}(t)~d^{3}\pmb{\mathrm{p}},
\label{eqn_methods_arb_j_int}
\end{equation}
where $g$ is the electron state degeneracy (e.g., number of Dirac cones in the first Brillouin zone) and the integral extends over the entire momentum space. When performing the integral in Eq. (\ref{eqn_methods_arb_j_int}), the $\mathcal{N}_{\pmb{\mathrm{p}}}$ terms in Eqs. (\ref{eq_jkx})-(\ref{eq_jkz}) must replaced by $\mathcal{N}_{\pmb{\mathrm{p}}}+1$, as part of a regularization procedure to prevent divergences at large momenta due to the assumption of an infinitely-extended Dirac cone~\cite{PhysRevB.48.11705,PhysRevB.82.201402,1367-2630-15-5-055021,PhysRevB.95.125408}. Eq. (\ref{eqn_methods_arb_j_int}) is equivalent to Eq. (\ref{eqn_arbitrary_J}) in the main text.

\textbf{Nonperturbative time-domain quantum simulations.} Equations (\ref{eqn_3D_NR}) and (\ref{eqn_3D_PR}) are discretized on a scaled momentum space grid ($q_{i} = v_{i}p_{i}$, where $i\in\{x,y,z\}$). We then numerically integrate them over time using the Dormand-Prince adaptive step solver (Boost C++ library). The single-electron currents are integrated over all momentum space using the trapezoidal rule.

While simulations of graphene and other 2D DSMs are manageable on a discretized momentum grid, the simulations quickly become intractible due to the additional dimension in 3D Dirac semimetals. When the incident laser is linearly polarized, we can exploit cylindrical symmetry by aligning the polarization axis parallel to $q_{z} = v_{z}p_{z}$, reducing the problem to an effectively 2D one. Equations~(\ref{eqn_3D_NR}) to (\ref{eqn_3D_PR}) then become
\begin{subequations}
\begin{align}
\label{eqn_3D_NRed}
\dot{\mathcal{N}}_{\pmb{\mathrm{p}}}(t) &= -2\dot{\theta}(t)\mathrm{Re}\Big{[} \mathit{\Gamma}_{\pmb{\mathrm{p}}}(t) \Big{]}  - \frac{1}{\tau}[\mathcal{N}_{\pmb{\mathrm{p}}}(t) - \mathcal{N}_{\pmb{\mathrm{p}}}(t_{0})]\\
\label{eqn_3D_GRed}
\dot{\mathit{\Gamma}}_{\pmb{\mathrm{p}}}(t) &= \frac{1}{2}\dot{\theta}(t)\mathcal{N}_{\pmb{\mathrm{p}}}(t) - \bigg{[}\frac{1}{\tau} + \frac{2\mathrm{i}\mathcal{E}(t)}{\hbar}\bigg{]} \mathit{\Gamma}_{\pmb{\mathrm{p}}}(t).
\end{align}
\end{subequations}
The time derivatives of the angular terms become $\dot{\phi} = 0$ and $\dot{\theta}(t) = ev_{z}E_{z}(t)q_{\rho}/\mathcal{E}(t)^{2}$, where $E_{z}(t)$ is the electric field component along $z$ and $q_{\rho}$ is the radial coordinate in the scaled momentum space. Note that the above equations and the initial conditions have no angular dependence, which is also the case for the single-electron current:
\begin{equation}
j_{\pmb{\mathrm{p}},z}(t) = -ev_{z}\Bigg{\{}\frac{[\mathcal{N}_{\pmb{\mathrm{p}}}(t)+1]q_{z} + 2 q_{\rho}\mathrm{Re}[ \mathit{\Gamma}_{\pmb{\mathrm{p}}}(t)]}{\sqrt{(q_{z}+ev_{z}A_{z})^{2} + q_{\rho}^{2}}}\Bigg{\}}.
\end{equation}
Since both the single-electron current and Eqs. (\ref{eqn_3D_NRed}) and (\ref{eqn_3D_GRed}) are rotationally-invariant about the axis of polarization, we have effectively reduced the 3D problem to a 2D one. Thus, for linearly polarized illumination, we compute the macroscopic current along $z$, integrated over all momentum-resolved contributions, as
\begin{equation}
J_{z}(t) \propto  2\pi\int^{+\infty}_{0}\int^{+\infty}_{-\infty}j_{\br{p},z}(t)q_{\rho}dq_{z}dq_{\rho}.
\end{equation}

\textbf{Modelling ultrafast THz pulses.} High intensity THz pulses from compact sources are frequently realized in the few-to-single cycle regime. To simulate a pulsed plane wave that does not contain spurious non-zero DC components, we chose a Poisson power spectrum, which accurately describes ultrafast pulses with durations down to single laser period~\cite{Winful2000}:
\begin{equation}
F(\omega) = 2\pi e^{\mathrm{i}\phi_{0}}\bigg{(} \frac{s}{\omega_{0}} \bigg{)}^{s+1}\frac{\omega^{s}\exp(-s\omega/\omega_{0})}{\Gamma(s+1)}\Theta(\omega).
\label{eqn_poisson_power_spectrum}
\end{equation}
Here, $\omega$ is angular frequency, $\phi_{0}$ is a phase constant, $s$ is a real, positive parameter that determines the pulse duration, $\omega_{0}$ is the peak angular frequency of the pulse, $\Gamma$ is the gamma function, and $\Theta$ is the Heaviside step function. In the narrow bandwidth limit (i.e., the many-cycle limit), Eq. (\ref{eqn_poisson_power_spectrum}) approaches a Gaussian spectrum of central angular frequency $\omega_{0}$. The electric field whose spectrum is given by Eq. (\ref{eqn_poisson_power_spectrum}) is
\begin{equation}
\pmb{\mathrm{E}}(t) = \mathrm{Re}\Bigg{\{} \pmb{\mathrm{E}}_{0}e^{\mathrm{i}\phi_{0}}\bigg{[}1 + \frac{\mathrm{i}\omega_{0}(t-t_{\mathrm{pk}})}{s} \bigg{]}^{-s-1} \Bigg{\}},
\end{equation}
where the peak amplitude is $\pmb{\mathrm{E}}_{0} = (E_{x,0},E_{y,0},E_{z,0})$  and $t_{\mathrm{pk}}$ is the time where the pulse reaches its peak. The intensity full width at half maximum (FWHM) $\tau_{\mathrm{FWHM}}$ is related to parameter $s$ via $\tau_{\mathrm{FWHM}} = 2s\sqrt{2\log(2)/(2s-1)}/\omega_{0}$.  Throughout this work, all pulse durations refer to the intensity FWHM.  For $\tau_{\mathrm{FWHM}} = 2$ ps, the corresponding shape factor is $s \approx 56.4$.  The vector potential $\pmb{\mathrm{A}}(t) = (A_{x}(t),A_{y}(t),A_{z}(t))$ at time $t$ is given by
\begin{equation}
\pmb{\mathrm{A}}(t) = -\int^{t}_{-\infty}\pmb{\mathrm{E}}(t')~dt'.
\end{equation}

\textbf{THz interaction with DSM thin film} To model the interaction of a linearly polarized driving THz pulse with a finite 3D DSM thin film at normal incidence, we employ a $1+1$D finite difference time domain (FDTD) routine which fully incorporates all intraband nonlinearities induced by the incident field. At the starting location $z = z_{\mathrm{in}}$, we have the input fields $\br{E}_{\mathrm{in}} = (E_{\mathrm{in},x},0,0)$ and $\br{H}_{\mathrm{in}} = (0,E_{\mathrm{in},x}/\eta_{0},0)$, where  $\eta_{0}$ is the free space impedence and $E_{\mathrm{in},x}$ is given by
\begin{equation}
E_{\mathrm{in},x}(z_{\mathrm{in}},t) = \mathrm{Re}\Bigg{\{}E_{x,0}e^{\mathrm{i}\phi_{0}}\bigg{[}1 - \frac{\mathrm{i}(k_{0}z' - \omega_{0} t)}{s} \bigg{]}^{-s-1}\Bigg{\}}.
\end{equation} 
The wavenumber is $k_{0} = \omega_{0}/c$. We define $z' = z_{\mathrm{in}} - z_{\mathrm{pk}}$, where $z_{\mathrm{pk}}$ is the initial pulse peak location.  Maxwell's equations are given by
\begin{subequations}
\begin{align}
\partial_{t}H_{y} &= -\frac{1}{\mu}\partial_{z}E_{x}\\
\epsilon_{0}\partial_{t}E_{x} &= -\partial_{z}H_{y} - J_{x}.
\end{align}
\end{subequations}
The total current is $J_{x} = J_{x}^{\mathrm{f}}+J_{x}^{\mathrm{p}}$, where $J_{x}^{\mathrm{f}}$ is the free current and $J_{x}^{\mathrm{P}} = \partial_{t}P_{x}$ is the current arising from the dielectric polarization $P_{x}$. To evaluate $J_{x}$, we use our analytical solutions for the intraband current, given by Eqs. (\ref{eqn_3D_intra_arb_pol_sub}) and (\ref{eqn_3D_intra_arb_pol_sup}). 

We discretize these equations on a Yee grid, which is staggered in both time and space for the evaluation of both the electric and magnetic fields.  Defining normalized time and spatial steps $\Delta(\omega_{0} t)$ and $\Delta(k_{0}z)$ respectively, we obtain
\begin{subequations}
\begin{align}
cB_{y}\rvert^{n+1/2}_{j+1/2} &= cB_{y}\rvert^{n-1/2}_{j+1/2} -\frac{\Delta (\omega_{0} t)}{\Delta (k_{0}z)}\Big{(} E_{x}\rvert^{n}_{j+1} - E_{x}\rvert^{n}_{j} \Big{)}\\
\begin{split}
E_{x}\rvert^{n+1}_{j} &= E_{x}\rvert^{n}_{j} - \frac{1}{\mu_{r}}\frac{\Delta (\omega_{0} t)}{\Delta (k_{0}z)}\Big{(} cB_{y}\rvert^{n+1/2}_{j+1/2} - cB_{y}\rvert^{n+1/2}_{j-1/2} \Big{)} \\
&\quad - \frac{\Delta (\omega_{0} t)}{\omega_{0} \epsilon_{0}}J_{x}\rvert^{n+1}_{j}.
\end{split}
\end{align}
\end{subequations}
We have assumed the material is linear and homogeneous in magnetic field response: $B_{y} = \mu_{0}\mu_{\mathrm{r}} H_{y}$ ($\mu_{\mathrm{r}} = 1$ through this work). The upper index, $n$, denotes the time step index. The lower index, $j$, denotes the spatial grid index. We also implement Mur absorbing boundary conditions~\cite{MurG} at both ends of our FDTD grid.

Note that $J_{x}\rvert^{n+1}_{j}$  makes this scheme implicit since the current depends on the field $E_{x}\rvert^{n+1}_{j}$ at the current time step. To obtain the correct $J_{x}\rvert^{n+1}_{j}$ and $E_{x}\rvert^{n+1}_{j}$, we employ a fixed-point interation method where $J_{x}\rvert^{n+1}_{j} = 0$ is used as an initial guess to compute $E_{x}\rvert^{n+1}_{j}$. We use this first pass solution to obtain a refined approximation of $J_{x}\rvert^{n+1}_{j}$. This procedure is iterated until the error between two consecutively refined values of $E_{x}\rvert^{n+1}_{j}$ are within a specified tolerance.  We varied this tolerance, the values of $\Delta(k_{0}z)$, the Courant number $\Delta(\omega_{0} t)/\Delta(k_{0}z)$, $z_{\mathrm{pk}}$, and simulation box width until convergence was achieved for the overall simulation. We assume free space on either side of the 3D DSM film, although this algorithm can be readily adapted to account for the presence of adjoining complex dispersive media (by incorporating the implementation in~\cite{Luebbers}), or adjoining nonlinear media of other kinds whose behavior can be captured by current $J_x$ as a function of $E_x$ like above.

We note that the use of a plane wave input pulse and a 1+1D FDTD -- as opposed to a focused laser pulse and a 3+1D FDTD -- ignores effects arising from beam diffraction and wavefront curvature. However, it should be noted that for a weakly focused laser pulse (even one of high field strengths), a plane wave pulse is a reasonable approximation.

Details of the implementation of the 1+1D FDTD algorithm for 2D DSMs (such as graphene) are given in SI Section VIII.

\textbf{Computing HHG spectra}  For an $x$-polarized plane wave pulse $E_{x}$ propagating in $z$ and therefore impinging on the DSM thin film at normal incidence, we can express the induced current density at position $z$ within the film as a linear combination of harmonic components:
\begin{equation}
J_{x}(z,t) = \mathrm{Re}\Bigg{[}\frac{1}{2\pi}\int \tilde{J}_{x}(z,\omega)e^{\mathrm{i}\omega t}d\omega\Bigg{]}.
\end{equation}
The energy radiated per unit solid angle $\Omega$ per unit angular frequency in the far-field is given by (derivation in SI Section IX)
\begin{equation}
\frac{d^{2}U}{d\omega d\Omega} = \frac{A^{2}}{8\pi^{3}\epsilon_{0}c}\big{(}\cos^{2}\phi\cos^{2}\theta + \sin^{2}\phi\big{)}\big{\lvert}  \tilde{F}(\theta,\omega) \big{\rvert}^{2}
\label{eqn_JperHzpersr}
\end{equation}
where $A$ is the area of the sample. We take $A = \pi R^{2}$, where $R = 1$ mm throughout this paper. $\tilde{F}(\theta,\omega)$ is given by the Fourier transform of
\begin{equation}
\begin{split}
F(\theta,t) &= \frac{1}{2\pi}\int\bigg{[}\mathrm{i}ke^{\mathrm{i}\omega t-\mathrm{i}kr} \frac{J_{1}(kR\sin\theta)}{kR\sin\theta}\\
&\qquad\qquad\times\int_{0}^{D} \tilde{J}_{x}(z',\omega)e^{\mathrm{i}kz'\cos\theta}dz'\bigg{]}d\omega.
\end{split}
\label{eqn_F_theta}
\end{equation}
The wavenumber corresponding to the angular frequency component $\omega$ is $k = \omega/c$, the first-order Bessel function is $J_{1}$, and the thickness of the DSM thin film is $D$ (we choose $D = 250$ nm throughout our work). 

For Fig.~\ref{fig_1}, we numerically evaluate the integral over $z'$ in Eq. (\ref{eqn_F_theta}) based on our FDTD results, followed by numerically integrating the intensity spectrum, given by Eq. (\ref{eqn_JperHzpersr}), over all solid angles in the forward emission direction to get the energy spectral density $dU/d\omega$. To obtain the spectrum in units of photons per 1\% bandwidth (BW), we divide $dU/d\omega$ by a factor  $100\hbar$. For Fig.~\ref{fig_2} and \ref{fig_3}, we present our results using Eq. (\ref{eqn_JperHzpersr}) evaluated at $\theta = 0$ (forward emission), and assume that the current is spatially uniform throughout the entire sample volume. 

\textbf{Computing energy efficiency} The energy conversion efficiency is defined as the ratio of the energy of the $N^{\mathrm{th}}$ harmonic, $U_{N}$, to the incident pulse energy $U_{\mathrm{in}}$.  The incident pulse energy is computed as
\begin{equation}
U_{\mathrm{in}} = \frac{A}{2\mu_{0} c}\int \big{\lvert}E_{\mathrm{in},x}(t)\big{\rvert}^{2} dt
\end{equation}
where $E_{\mathrm{in},x}(t)$ is the temporal profile of the incident driving electric field and $A$ is the area of the sample as defined in Eq.~(\ref{eqn_JperHzpersr}). To get the energy conversion efficiency of the $N^{\mathrm{th}}$ harmonic, we integrate the energy spectral density $dU/d\omega$ over the frequency domain from $(N-1)\omega_{0}$ to $(N+1)\omega_{0}$.

\centerline{}
\begin{acknowledgements}
\textbf{Acknowledgements} We acknowledge the National Supercomputing Center (NSCC) Singapore for the use of their computing resources. L.J.W. acknowledges the support of the Agency for Science, Technology and Research (A*STAR) Advanced Manufacturing and Engineering Young Individual Research Grant (A1984c0043); and the Nanyang Assistant Professorship Start-up Grant.  L.K.A. and J.L. acknowledge funding from A*STAR IRG (A1783c0011), MOE PhD RSS, and USA ONRG grant (N62909-19-1-2047). I.K. acknowledges the support of the Azrieli Faculty Fellowship, the Israel Science Foundation grant no. 3334/19 and 831/19, and the ERC starting grant NanoEP 851780 from the European Research Council. F.J.G.d.A acknowledges funding from ERC (Advanced Grant No. 789104-eNANO) and Spanish MINECO (Grants No. MAT2017-88492-R and No. SEV2015-0522) .
\centerline{}
\textbf{Author contributions} All authors made critical contributions in conceiving the study, analyzing the results, and writing the manuscript.
\centerline{}
\textbf{Additional Information} Supplementary Information accompanies this paper at XXXXX
\centerline{}
\textbf{Competeting financial interests} The authors declare no competing financial interests
\end{acknowledgements}

\bibliographystyle{apsrev4-1} 
\bibliography{main_3D_dirac_HHG}  

\end{document}